\documentclass[a4paper,12pt]{article}

\usepackage[T2A]{fontenc}
\usepackage{amsmath,amssymb,amsthm}
\usepackage{dsfont}
\usepackage{cite}
\usepackage[unicode,linktocpage=true,plainpages=false,pdfpagelabels=false]{hyperref}

\newcommand{\dd}{\partial}
\newcommand{\de}{\delta}
\newcommand{\m}{\mu}
\newcommand{\n}{\nu}
\newcommand{\ka}{\varkappa}
\newcommand{\si}{\sigma}
\newcommand{\al}{\alpha}
\newcommand{\be}{\beta}
\newcommand{\Ga}{\Gamma}
\newcommand{\ep}{\varepsilon}
\begin{document}

\title{Investigation of Faddeev variant of embedding theory}
\author{
S.~S.~Kuptsov\thanks{E-mail: s1t2a3s4@yandex.ru},
S.~A.~Paston\thanks{E-mail: pastonsergey@gmail.com}
\\
{\it Saint Petersburg State University, Saint Petersburg, Russia}
}
\date{\vskip 15mm}
\maketitle

\begin{abstract}
Faddeev variant of embedding theory is an example of using the embedding approach for the description of gravity.
In the original form of the embedding approach, the gravity is described by an embedding function of a four-dimensional surface representing our spacetime. In Faddeev variant, the independent variable is a non-square vielbein, which is a derivative of embedding function in embedding theory. We study the possibility of the existence of extra solutions in Faddeev variant, which makes this theory non-equivalent to GR. To separate the degrees of freedom corresponding to extra matter, we propose a formulation of this theory as GR with an additional contribution to the action. We analyze the equations of motion for a specific class of solutions corresponding to a weak gravitational field. We construct a simple exact solution corresponding to arbitrary matter and nontrivial torsion, which is an extra solution in Faddeev variant in the absence of real matter.
\end{abstract}

\newpage
	
\section{Introduction}
Embedding theory (or "embedding gravity"{}) was proposed in \cite{regge} as a modified gravity inspired by string theory. In this approach we consider our spacetime as 4D surface in 10D Minkowski ambient space described by embedding function $y^a(x^\m)$ (here and hereafter $\m,\n,\ldots=0,1,2,3$; $a,b,\ldots=0,\ldots,9$). After the appearance of this approach, its aspects were discussed by many authors, in particular, within the context of quantization of gravity and explanation of dark matter and energy; see, e.g.
\cite{deser,tapia,estabrook1999,davkar,statja18,rojas09,statja25}.
Faddeev proposed \cite{faddeev} a variant of embedding theory, in which the role of an independent variable is played by non-square vielbein $e^a_\mu$. In the original form of embedding theory, this quantity can be expressed in terms of embedding function:
$e^a_\mu=\dd_\m y^a$, which imposes serious restrictions on $e^a_\mu$. On the contrary, these restrictions are absent in Faddeev variant.
Both metric and connection in Faddeev variant can be expressed using $e^a_\mu$ in the following form:
\begin{align}
	g_{\mu\nu} = e_\mu^a e_\nu^b \eta_{a b},\qquad
	\Gamma^\alpha_{\mu\nu} = e^\alpha_b \partial_\mu e_\nu^b, \label{con}
\end{align}
where $\eta_{a b}$ is a flat metric of ambient space (for which connectivity and torsion are equal to zero), $e^\alpha_b$ is a reverse vielbein, $e^\alpha_b = \eta_{ab} g^{\alpha\beta} e_\beta^a$. The expression for metric has the same form as in the tetrad formalism. They only differ in number of values which index $a$ can take, which justifies the naming \emph{non-square vielbein} for $e^a_\mu$.

As in the usual formulation, the gravitational action in Faddeev variant is chosen to be Einstein-Hilbert one. It can be shown \cite{faddeev} that the equations of motion for this theory have the form (we restrict ourselves to a vacuum case):
\begin{align}
	G_{\mu\nu} =  0,\qquad
	b_{\mu\nu}^a S_{\alpha\beta\delta} \varepsilon^{\mu\beta\delta\xi} \varepsilon^{\sigma\nu\alpha\zeta} g_{\xi\zeta} = 0,
	\label{eqs}
\end{align}
where $S^\alpha{}_{\beta\delta}=\Ga^\alpha_{\beta\delta}-\Ga^\alpha_{\delta\beta}$ is a torsion tensor and $G_{\mu\nu}$ is the Einstein tensor
(generally speaking, it is non-symmetric due to the presence of the torsion, so the first formula of \eqref{eqs} contains 16 equations rather than 10),
$b^a_{\m\nu}$ is a second fundamental form of the surface, which can be written using the projector $\Pi_{\bot}^{ab}$ to the subspace orthogonal to the vielbein:
\begin{equation}
	b^a_{\mu\nu} = {\Pi_{\bot}}^a_b\, \partial_\mu e_\nu^b, \qquad \Pi_{\bot}^{ab} = \eta^{ab} - e^a_\mu e^b_\nu g^{\mu\nu}.
	\label{2osn}
\end{equation}
Taking such orthogonality into account, we see that the second formula of \eqref{eqs}, which has free indices  $a$ and $\si$, contains $6\cdot 4=24$ equations.

It is clear that the second equation of \eqref{eqs} can be satisfied by the assumption of vanishing torsion \cite{faddeev}. In this case, the connection defined by \eqref{con} turns out to be consistent with metric (as in pseudo-Riemannian geometry), so the first equation of \eqref{eqs} coincides with the vacuum Einstein equation. The solutions with zero torsion, therefore, resemble the solutions of GR. However, the question remains about the existence of solutions of \eqref{eqs} with nonzero torsion. Section 3 of the present work is devoted to the discussion of this question.

If such solutions exist, then the corresponding metric would not be a solution of the vacuum Einstein equations. However, one can assume that it satisfies Einstein equations with some nonzero r.h.s. (note that the presence of torsion could not affect the observables directly in case if a matter interacts only with the metric). It would mean that Faddeev variant of embedding theory is a modified gravity theory, which is more general than GR since, besides "Einsteinian" solutions, it also contains "extra" solutions, which can be interpreted as a presence of some \emph{extra} matter. Such modifications, among which are embedding gravity itself  \cite{pavsic85} and recently popular mimetic gravity \cite{mukhanov}, could be of use in the search for an explanation of the mystery of dark matter \cite{statja68,mimetic-review17}. In Section 2, we show how one can reformulate  Faddeev variant of embedding theory as GR with the extra matter. To do that, we write down the action of the theory, which is a sum of Einstein-Hilbert one and the extra matter term.

\section{The action of the theory in the form of GR with extra matter}
To simplify the study of properties of extra matter, one can rewrite the considered theory as a GR with a certain additional contribution to the action, which describes this matter. Such trick was proposed in \cite{Golovnev201439} for mimetic gravity and was subsequently used in many works; for embedding gravity, it was proposed in \cite{statja48,statja51} and was used in \cite{statja67,statja68} for the analysis of the properties of dark matter.
Here is one of possible forms of additional matter term suitable for Faddeev approach:
\begin{equation}
	S
	= -\frac{1}{2\ka} \int d^4x \sqrt{-g} \left(
	{R}(g) + \left[K^\alpha{}_{\alpha\beta}(e) K^\beta{}_{\nu\mu}(e)\right]_{\alpha\nu} g^{\mu\nu} +
	\tau^{\mu\nu}\left(g_{\mu\nu} - e_\mu^a e_{a\nu}\right)\right),
	\label{S}
\end{equation}
where $[\ldots]_{\alpha\nu}$ ($\{\ldots\}_{\alpha\nu}$) here and hereafter means antisymmetrization (symmetrization) of bracketed expression w.r.t. upper or lower indices mentioned (without the multiplier $1/2$). In  \eqref{S} $g_{\m\n}$ is an independent metric, ${R}(g)$ is the scalar curvature expressed through it;\\
$K_{\mu\alpha\beta} = \left(S_{\mu\alpha\beta} - S_{\alpha\beta\mu} - S_{\beta\alpha\mu}\right)/2$ is the contorsion tensor, whose first index can be raised and lowered using a metric constructed out of vielbein $\widetilde{g}_{\mu\nu}\equiv e_\mu^a e_{a\nu}$.
The set of independent variables in the considered action consists of  $g_{\m\n}$, $e_\mu^a$, and symmetric tensor $\tau^{\mu\nu}$, which plays the role of the Lagrange multiplier.
It can be shown that after the substitution of $g_{\mu\nu}$ by $\widetilde{g}_{\mu\nu}$ the first two terms in the action \eqref{S} match the expression for the scalar curvature written through metric and connection \eqref{con} up to a full covariant derivative. Therefore they reproduce the action of Faddeev variant of embedding theory with $g_{\mu\nu} = \widetilde{g}_{\mu\nu}$ up to a surface term.

Let us obtain the equations of motion corresponding to the action \eqref{S} by varying it w.r.t. all independent variables. The variation of \eqref{S} w.r.t. non-square vielbein $e_\mu^a$ has a part which is longitudinal w.r.t. vielbein as well as transverse one. The symmetric part of longitudinal equations allows to express $\tau^{\mu\nu}$ in terms of other variables and exclude it from the equations of motion. As a result, the equations of motion take the form of Einstein equations with energy-matter tensor (EMT) of extra matter, together with the equations of motion of extra matter itself described by non-square vielbein $e^a_\mu$. The Einstein equations can be written as
\begin{align}
	G_{\mu\nu}(g) = \ka\, T_{\m\n}(e,g),
	\label{eqs_einst}
\end{align}
where $G_{\mu\nu}(g)$ is a standard pseudo-Riemannian expression for Einstein tensor defined by metric, and the EMT has the form
\begin{equation}
	T_{\m\n}(e,g) = -\frac{1}{2\ka}\left\{\left[K^\alpha{}_{\alpha\beta} K^\beta{}_{\nu\mu} + D_\alpha K^\alpha{}_{\nu\mu}\right]_{\alpha\nu} -
	\frac{1}{2} g_{\mu\nu} \left[K^\alpha{}_{\alpha\beta} K^\beta{}_{\gamma\delta} + D_\alpha K^\alpha{}_{\gamma\delta}\right]_{\alpha\gamma} g^{\gamma\delta}\right\}_{\mu\nu}.		 \label{T}
\end{equation}
Here and hereinafter $D_\alpha$ means covariant derivative which is constructed using the Levi-Civita connection from Riemannian geometry. Note that the gravitational constant $\ka$ is introduced here to ensure the usual form of Einstein equations. The equations of motion of extra matter have the form
\begin{align}
&\phantom{[}b_{\mu\nu}^c S_{\alpha\beta\delta} \varepsilon^{\mu\beta\delta\xi} \varepsilon^{\sigma\nu\alpha\zeta} g_{\xi\zeta} = 0, 		\label{bS=0}\\
&\left[K^\alpha{}_{\alpha\beta} K^\beta{}_{\nu\mu}-K^\alpha{}_{\nu\beta} K^\beta{}_{\alpha\mu} + D_\alpha K^\alpha{}_{\nu\mu}-D_\nu K^\alpha{}_{\alpha\mu}\right]_{\mu\nu} = 0, \label{eqs_asym}\\
&\phantom{[} e_\mu^a e_{a\nu} = g_{\mu\nu},		\label{g=g}
\end{align}
where \eqref{bS=0} is the transverse part of the equation obtained by variation of the action w.r.t. non-square vielbein $e_\si^c$,
\eqref{eqs_asym} is an asymmetric part of the corresponding longitudinal equation, and
\eqref{g=g} appears as a result of variation w.r.t. Lagrange multiplier $\tau^{\mu\nu}$.

When the equations of the theory are written in the form  \eqref{eqs_einst}-\eqref{g=g},
the metric degrees of freedom (corresponding to gravitational field in the GR framework) and the extra matter ones turn out to be separated, which could help to analyze the theory in the context of its interpretation as GR with the extra matter. The equations  \eqref{bS=0}-\eqref{g=g} can be considered as the equations of motion of the extra matter in a given gravitational field, and the Einstein equation  \eqref{eqs_einst} as the equation which determines the metric (as in the usual approach). The direct calculation shows that after the exclusion of independent metric the set of equations  \eqref{eqs_einst} and \eqref{bS=0}-\eqref{g=g} is equivalent to the system \eqref{eqs}, i.e. \eqref{S} indeed describes Faddeev variant of embedding theory.

First of all, in the context of the interpretation of extra matter as dark matter, it is interesting to consider the weak gravitational field limit when the independent metric has the form $g_{\m\n}=\eta_{\m\n}+h_{\m\n}$ with small $h_{\m\n}$. In the zeroth-order, one can assume $g_{\m\n}=\eta_{\m\n}$ while solving the equations of motion of extra matter. Then one can consider corrections to the obtained zeroth order solution and analyze the EMT \eqref{T} behavior. The corresponding full analysis is beyond the scope of the present paper, in which we will restrict ourselves to the study of the possibility to use a specific ansatz in search for "extra" solutions corresponding to nonzero torsion (and thus nonzero extra matter EMT). We will search for corresponding solution in the weak gravitational field approximation, using the original equations \eqref{eqs} rather than \eqref{eqs_einst}-\eqref{g=g}.

\section{The search for extra solutions in the weak gravitational field limit}
Let us search for solutions of such class for which the metric constructed out of vielbein
is close to $\eta_{\mu\nu}$. We want to parametrize the vielbein in such a way that 10 parameters, which we denote by $h_{\mu\nu}$, would determine the leading contribution to the quantity $g_{\mu\nu}-\eta_{\mu\nu}$. All other degrees of freedom do not affect the metric directly, so we consider them as degrees of freedom of extra matter. Then 40 equations \eqref{eqs} can be considered as 30 equations of motion of extra matter together with 10 equations of its coupling with metric.

Let us represent the vielbein as an asymptotic series, for which the zeroth order approximation $\overset{_{(0)}}{e}\vphantom{e}^a_\mu$ ensures $g_{\mu\nu}=\eta_{\mu\nu}$ at the leading order as well as satisfies the equations \eqref{eqs}. It can be shown that $\overset{_{(0)}}{e}\vphantom{e}^a_\mu = \Lambda_\mu{}^\gamma \delta^a_\gamma$ with an arbitrary pseudo-orthogonal matrix
 $\Lambda_\mu{}^\gamma(x)$ will work. However, we restrict ourselves to the simplest case  $\overset{_{(0)}}{e}\vphantom{e}^a_\mu = \delta^a_\mu$, when the vielbein is tangent to the first four axes of ambient space coordinates. Let us introduce the indices  $I,J,\ldots=4,\dots,9$, so $a$ runs over $\{\mu,I\}$. Then the problem of separation of metric degrees of freedom can be solved by the following ansatz:
\begin{equation}
e^a_\mu = \sum_{k=0}^\infty \varepsilon^k \overset{_{(k)}}{e}\vphantom{e}^a_\mu =  \delta^a_\mu +
\varepsilon \left(\frac{1}{2} h_\mu{}^\gamma + \Omega_\mu{}^\gamma \right) \delta^a_\gamma +
\varepsilon U_\mu{}^I \delta_I^a +
O(\varepsilon^2),
\label{anzatz}
\end{equation}
where $\varepsilon$ is a small asymptotic parameter of our series, the symmetric tensor
$h_{\mu\nu}$ is responsible for the leading order correction of a flat metric, and antisymmetric  $\Omega_{\mu\gamma}$ with quantity $U_\mu{}^I$ corresponds to the extra matter
(here and hereinafter we assume that the indices $\mu,\nu,\ldots$ raise and lower using  $\eta_{\mu\nu}$).

It is expected that after the substitution of \eqref{anzatz} in the equations the first nontrivial order will determine $h_{\m\n}$, $\Omega_{\m\n}$ and $U_\m{}^I$, the next one --- $\overset{_{(2)}}{e}\vphantom{e}^a_\m$ and so on. We restrict our consideration of all expressions to the leading order. Note that the ansatz \eqref{anzatz} is not the most general one in such a sense that all components of the leading correction of $\delta^a_\mu$ have the same order of magnitude here. One can surely consider an ansatz in which $\Omega$ and $h$ have the same order as $U^2$. In this case, the way \eqref{anzatz} depends on $h_{\m\n}$ needs to be changed considering the contributions of $U_\mu{}^I U_{\nu I}$ to the metric.

Let us calculate all quantities which are present in the equations \eqref{eqs} with sufficient accuracy.
We start with the metric:
\begin{equation}
	g_{\mu\nu} = \eta_{\mu\nu} +
	\varepsilon h_{\mu\nu} +
	O(\varepsilon^2).
\end{equation}
For the longitudinal projector $\Pi^{a b} = e^a_\mu e^b_\nu g^{\mu\nu}$ some cancellations occur, which result in the following form of the transverse projector which is of our interest:
\begin{equation}
\Pi_{\bot}^{ab}=
\delta^a_I \delta^b_J \delta^{I J} -
\ep \left( \eta^{a\m} \de^b_I U_\m{}^I +\eta^{b\m} \de^a_I U_\m{}^I\right)+
O(\varepsilon^2).
\label{Pi}
\end{equation}
Using the expression for the Ricci tensor in the approach with independent non-square vielbein and taking into account that
$\Pi_{\bot}^{a\m}$ in the zeroth-order is equal to zero, we obtain that
\begin{equation}
	R_{\mu\nu} = g^{\al\be}\Pi_{\bot}^{ab}\left[\partial_\alpha e_{a\be} \partial_\nu e_{b \mu} \right]_{\alpha\nu} =
	\varepsilon^2 \left[\partial_\alpha U^{\alpha I} \partial_\nu U_{\mu I} \right]_{\alpha\nu} +
	O(\varepsilon^3),
	\label{rich}
\end{equation}
so the variables $h_{\m\n}$ and $\Omega_{\m\n}$ do not contribute to the leading order of the Ricci tensor, and therefore to the leading order of the first equations of \eqref{eqs}. The same situation occurs with the second fundamental form \eqref{2osn}:
\begin{equation}
	b^a_{\mu\nu} = \varepsilon \delta_I^a \partial_\mu U_\nu{}^I +
	O(\varepsilon^2).
\label{sp1}
\end{equation}
Therefore the only possibility for the components  $h_{\m\n}$ and $\Omega_{\m\n}$ to contribute to the leading order of equations \eqref{eqs} is through torsion constructed from the connection
\begin{equation}
	\Gamma_{\alpha\beta\delta} = \varepsilon \partial_\beta \left(\frac{1}{2} h_{\delta\alpha} +
	\Omega_{\delta\alpha}\right) + O(\varepsilon^2).
\end{equation}

Now everything is ready for us to write the equations \eqref{eqs} in the first nontrivial order. We restrict ourselves to a vacuum case, when the first equation $G_{\mu\nu}=0$ means $R_{\mu\nu}=0$. Note again that we are looking for solutions with non-zero torsion, so it does not follow from \eqref{eqs} that $G_{\mu\nu}\left( g \right) = 0$, and we expect a non-zero $T_{\mu\nu}$ in  \eqref{eqs_einst}.  The resulting equations have the form:
\begin{equation}
\left[\partial_\alpha U^{\alpha I} \partial_\nu U_{\mu I} \right]_{\alpha\nu} = 0,\qquad
\eta_{\xi\zeta}\varepsilon^{\mu\beta\delta\xi} \varepsilon^{\sigma\nu\alpha\zeta}
\left(\partial_\mu U_\nu{}^I\right)
\left[\partial_\beta \left(\frac{1}{2} h_{\delta\alpha} +
\Omega_{\delta\alpha}\right)\right]_{\be\de}  = 0.
\label{eqs_anzatz1}
\end{equation}
The first of these equations are 16 equations governing the behavior of 24 components of $U_\m{}^I$, and the second ones are 24 equations for  $h_{\m\n}$, $\Omega_{\m\n}$ and the remaining 8 components of $U_\m{}^I$.
Since the part of the first equations that is symmetric w.r.t. free indices $\m,\n$ corresponds to Einstein equations \eqref{eqs_einst} in the context of the consideration of the theory as GR with the extra matter, it is tempting to find the correction of the flat metric $h_{\m\n}$ from them. However, as can be seen, the ansatz we used prohibits that; furthermore, it can be shown that the situation would remain the same even if one makes $U_\m{}^I$ smaller relative to $h_{\m\n}$ and $\Omega_{\m\n}$ and takes higher orders into account (it is connected to the possibility to rewrite the Ricci tensor in terms of derivatives of the projector $\Pi^{a b}$).

Note that the second equation
\eqref{eqs_anzatz1} can be treated as an action of $24\times24$ matrix (dependent on $U_\m{}^I$) on the vector constructed out of 24 torsion components (it is the antisymmetrization result in this formula up to a multiplier $\ep$). If one uses the arbitrariness in the choice of $U_\m{}^I$ remaining after the solution of the first equations \eqref{eqs_anzatz1}, one can require the vanishing of the determinant of this matrix. Then among the solutions might appear the ones that correspond to nonzero torsion, but this question requires an additional study.

The equations \eqref{eqs_anzatz1} admit the most simple solution $U_\m{}^I = 0$. Generally speaking, this solution is prohibited when one uses the ansatz \eqref{anzatz} and leading order approximations since the used order disappears in this solution. However, it turns out that this solution satisfies exact equations \eqref{eqs} also, if $\overset{_{(k)}}{e}\vphantom{e}^a_\mu$ at $k>1$ together with $U_\m{}^I$ are vanished in \eqref{anzatz}, i.e. all higher-order corrections are absent. In this case, $\ep$ is not necessarily small, i.e., one can choose the non-square vielbein in the form
\begin{equation}
	e^a_\mu = V_\mu{}^\gamma \delta^a_\gamma,	\label{reshen}
\end{equation}
where  $V_\mu{}^\gamma(x)$ is an arbitrary non-degenerate matrix. It can be checked that \eqref{reshen} is an exact solution of \eqref{eqs}. This fact has a clear geometric meaning: the projector $\Pi^{a b}$ corresponding to \eqref{reshen} turns out to be the same that the one that corresponds to $e^a_\mu = \delta^a_\mu$, i.e., a constant one. Since it can be shown that in the theory with independent non-square vielbein, the second fundamental form  $b^a_{\mu\nu}$ and the curvature tensor turn out to be proportional to derivatives of the projector, the equations \eqref{eqs} turn out to be satisfied.

Therefore, we have a class of extra solutions of the equations  \eqref{eqs}, in which the metric and the connection depend on a completely arbitrary function:
\begin{equation}
	g_{\mu\nu} = V_\mu{}^\alpha V_\nu{}^\beta \eta_{\alpha\beta}, \qquad \Gamma^\alpha{}_{\mu\nu} = V^{-1}{}_\gamma{}^\alpha \partial_\mu V_\nu{}^{\gamma}.		\label{g(V)}
\end{equation}
In this case, the torsion is nonzero, and the metric satisfies the Einstein equations
 \eqref{eqs_einst} with EMT contribution \eqref{T}.
The extra matter EMT itself is practically arbitrary; it can be expressed through expression corresponding to pseudo-Riemannian expression for Einstein tensor by substituting the metric in the form \eqref{g(V)}. Note that this exact solution
 \eqref{reshen} is valid only in the considered case of absent real matter.

\section{Conclusion}
We study the possibility of the existence of extra solutions (i.e., corresponding to the nonzero value of Einstein tensor constructed out of metric in the absence of real matter)
for Faddeev variant of embedding theory.
We have shown that this theory can be reformulated as GR with an additional contribution to the action corresponding to some fictitious \emph{extra} matter \eqref{S}. This matter does not represent any real particles or fields. It is a part of the gravitational degrees of freedom described by the non-square vielbein in the Faddeev approach.

It turns out (in the weak gravitational field approximation and with the assumption that the vielbein is close to $\delta^a_\mu$) that the metric degrees of freedom  $h_{\m\n}$ is explicitly separated from the degrees of freedom of extra matter $\Omega_{\m\n}$ and $U_\mu{}^I$ \eqref{anzatz}. Unexpectedly, the part of the equations of motion in the leading order, which corresponds to Einstein equations, turns out to be independent of $h_{\m\n}$. The obtained nonlinear equations require further study, in particular, within the context of the existence of their nontrivial solutions (corresponding to $U_\mu{}^I\ne0$).
However, we can present a simple exact solution \eqref{reshen} of the equations of motion for this theory. It means that in the vacuum case of Faddeev variant of embedding theory, there is at least some class of extra solutions with nonzero torsion and EMT of extra matter.

To study more diverse variants of the extra matter behavior, one can search for other extra solutions using \eqref{eqs_anzatz1} or use an ansatz for vielbein that differs from \eqref{anzatz}, which would be less trivial from a geometrical point of view. The properties of extra matter obtained in this way must be compared with known observational properties of dark matter and dark energy to determine whether Faddeev variant of embedding theory can help to solve the mystery of dark matter and dark energy.

{\bf Acknowledgements.}
The work is supported by RFBR Grant No.~20-01-00081.

%\newcommand{\eprint}[1]{\href{http://arxiv.org/abs/#1}{\texttt{#1}}}
%\bibliographystyle{../../my3}
%\bibliography{../../paston-grav-e}

\begin{thebibliography}{10}
\newcommand{\enquote}[1]{``#1''}
\providecommand{\url}[1]{\texttt{#1}}
\providecommand{\urlprefix}{URL }
\expandafter\ifx\csname urlstyle\endcsname\relax
  \providecommand{\doi}[1]{doi:\discretionary{}{}{}#1}\else
  \providecommand{\doi}{doi:\discretionary{}{}{}\begingroup
  \urlstyle{rm}\Url}\fi
\providecommand{\eprint}[1]{\href{http://arxiv.org/abs/#1}{\texttt{#1}}}

\bibitem{regge}
T.~Regge, C.~Teitelboim, \enquote{General relativity \`a la string: a progress
  report}, in \emph{Proceedings of the First Marcel Grossmann Meeting, Trieste,
  Italy, 1975}, edited by R.~Ruffini, 77--88, North Holland, Amsterdam, 1977,
  \eprint{arXiv:1612.05256}.

\bibitem{deser}
S.~Deser, F.~A.~E. Pirani, D.~C. Robinson, \enquote{New embedding model of
  general relativity},
  \href{http://dx.doi.org/10.1103/PhysRevD.14.3301}{\emph{Phys. Rev. D}},
  \textbf{14}: 12 (1976), 3301--3303.

\bibitem{tapia}
V.~Tapia, \enquote{Gravitation a la string},
  \href{http://dx.doi.org/10.1088/0264-9381/6/3/003}{\emph{Class. Quant.
  Grav.}}, \textbf{6} (1989), L49.

\bibitem{estabrook1999}
F.~B. Estabrook, R.~S. Robinson, H.~R Wahlquist, \enquote{Constraint-free
  theories of gravitation},
  \href{http://dx.doi.org/10.1088/0264-9381/16/3/019}{\emph{Class. Quant.
  Grav.}}, \textbf{16} (1999), 911--918.

\bibitem{davkar}
D.~Karasik, A.~Davidson, \enquote{Geodetic Brane Gravity},
  \href{http://dx.doi.org/10.1103/PhysRevD.67.064012}{\emph{Phys. Rev. D}},
  \textbf{67} (2003), 064012, \eprint{arXiv:gr-qc/0207061}.

\bibitem{statja18}
S.~A. Paston, V.~A. Franke, \enquote{Canonical formulation of the embedded
  theory of gravity equivalent to Einstein's general relativity},
  \href{http://dx.doi.org/10.1007/s11232-007-0134-9}{\emph{Theor. Math.
  Phys.}}, \textbf{153}: 2 (2007), 1582--1596, \eprint{arXiv:0711.0576}.

\bibitem{rojas09}
R.~Cordero, A.~Molgado, E.~Rojas, \enquote{Ostrogradski approach for the
  Regge-Teitelboim type cosmology},
  \href{http://dx.doi.org/10.1103/PhysRevD.79.024024}{\emph{Phys. Rev. D}},
  \textbf{79} (2009), 024024, \eprint{arXiv:0901.1938}.

\bibitem{statja25}
S.~A. Paston, \enquote{Gravity as a field theory in flat space-time},
  \href{http://dx.doi.org/10.1007/s11232-011-0138-3}{\emph{Theor. Math.
  Phys.}}, \textbf{169}: 2 (2011), 1611--1619, \eprint{arXiv:1111.1104}.

\bibitem{faddeev}
L.~D. Faddeev, \enquote{New dynamical variables in Einstein's theory of
  gravity}, \href{http://dx.doi.org/10.1007/s11232-011-0023-0}{\emph{Theor.
  Math. Phys.}}, \textbf{166}: 3 (2011), 279--290,
  \eprint{arXiv:0906.4639}, \eprint{arXiv:0911.0282}, \eprint{arXiv:1003.2311}.

\bibitem{pavsic85}
M.~Pavsic, \enquote{On The Quantization Of Gravity By Embedding Space-Time In A
  Higher Dimensional Space},
  \href{http://dx.doi.org/10.1088/0264-9381/2/6/012}{\emph{Class. Quant.
  Grav.}}, \textbf{2} (1985), 869, \eprint{arXiv:1403.6316}.

\bibitem{mukhanov}
A.~H. Chamseddine, V.~Mukhanov, \enquote{Mimetic dark matter},
  \href{http://dx.doi.org/10.1007/JHEP11(2013)135}{\emph{Journal of High Energy
  Physics}}, \textbf{2013}: 11 (2013), 135, \eprint{arXiv:1308.5410}.

\bibitem{statja68}
S.~A. Paston, \enquote{Non-relativistic limit of embedding gravity as General
  Relativity with dark matter},
  \href{http://dx.doi.org/10.3390/universe6100163}{\emph{Universe}},
  \textbf{6}: 10 (2020), 163, \eprint{arXiv:2009.06950}.

\bibitem{mimetic-review17}
L.~Sebastiani, S.~Vagnozzi, R.~Myrzakulov, \enquote{Mimetic Gravity: A Review
  of Recent Developments and Applications to Cosmology and Astrophysics},
  \href{http://dx.doi.org/10.1155/2017/3156915}{\emph{Advances in High Energy
  Physics}}, \textbf{2017} (2017), 3156915, \eprint{arXiv:1612.08661}.

\bibitem{Golovnev201439}
A.~Golovnev, \enquote{On the recently proposed mimetic Dark Matter},
  \href{http://dx.doi.org/10.1016/j.physletb.2013.11.026}{\emph{Physics Letters
  B}}, \textbf{728} (2014), 39 -- 40, \eprint{arXiv:1310.2790}.

\bibitem{statja48}
S.~A. Paston, \enquote{Forms of action for perfect fluid in general relativity
  and mimetic gravity},
  \href{http://dx.doi.org/10.1103/PhysRevD.96.084059}{\emph{Phys. Rev. D}},
  \textbf{96}: 8 (2017), 084059, \eprint{arXiv:1708.03944}.

\bibitem{statja51}
S.~A. Paston, A.~A. Sheykin, \enquote{Embedding theory as new geometrical
  mimetic gravity},
  \href{http://dx.doi.org/10.1140/epjc/s10052-018-6474-9}{\emph{The European
  Physical Journal C}}, \textbf{78}: 12 (2018), 989, \eprint{arXiv:1806.10902}.

\bibitem{statja67}
S.~A. Paston, \enquote{Dark matter from non-relativistic embedding gravity},
  \href{http://dx.doi.org/10.1142/S0217732321501017}{\emph{Modern Physics
  Letters A}}, \textbf{36}: 15 (2021), 2150101, \eprint{arXiv:2006.09026}.

\end{thebibliography}
%\end{document}

\end{document}